\documentclass[aps,pra,twocolumn,superscriptaddress,showpacs]{revtex4}
\usepackage{graphicx}
\usepackage{amsmath}
\usepackage{epstopdf}
\newcommand{\eq}[1]{Eq.(\ref{#1})}
\newcommand{\fig}[1]{Fig.~\ref{#1}}
\begin{document}

\title{Many-body theory of excitation dynamics in an ultracold Rydberg gas}

\author{C.\ Ates}
\affiliation{Max Planck Institute for the Physics of Complex Systems,
N{\"o}thnitzer Stra{\ss}e 38, D-01187 Dresden, Germany}
\author{T.\ Pohl}
\affiliation{ITAMP, Harvard-Smithsonian Center for Astrophysics, 60 Garden
Street, MS14, Cambridge, MA 02138, USA}
\author{T.\ Pattard\footnote{present address: APS Editorial Office, 1 Research
Road, Ridge, NY 11961, USA}}
\author{J.M.\ Rost}
\affiliation{Max Planck Institute for the Physics of Complex Systems,
N{\"o}thnitzer Stra{\ss}e 38, D-01187 Dresden, Germany}

\date{\today}

\begin{abstract}
We develop a theoretical approach for the dynamics of Rydberg excitations in ultracold gases, with a realistically large number of atoms.
We rely on the reduction of the single-atom Bloch equations to rate equations, which is possible under various experimentally relevant conditions. 
Here, we explicitly refer to a two-step excitation-scheme. We discuss the conditions under which our approach is valid by comparing the results  with the solution of the exact quantum master equation for two interacting atoms. 
Concerning the emergence of an excitation blockade in a Rydberg gas, our results are in qualitative agreement with experiment.  Possible sources of quantitative discrepancy are carefully examined. Based on the two-step excitation scheme, we predict the occurrence of an antiblockade effect and propose possible ways to detect this excitation enhancement experimentally in an optical lattice as well as in the gas phase.  
\end{abstract}


\maketitle

\section{Introduction} \label{intro}
The possibility to routinely create samples of ultracold gases in the
$\mu$-Kelvin regime has opened a new avenue to the investigation of interacting
many-particle systems.  At such temperatures, the thermal velocities
of the atoms are low enough that the atoms move a negligible distance
over the duration of the experiment.  Hence, (thermal) collisions are
not relevant and it is possible to study quasi-static interactions
between the particles.

For densities of a dilute ultracold but non-degenerate gas
typical for atoms in magneto-optical traps, the interaction between
ground state atoms is very weak.  Rydberg atoms, on the other hand, can
strongly interact among each other, even in a dilute gas, due to their
large polarizability which scales with the principal quantum number
$n$ as $n^7$.  This scaling allows an accurate control over their
interactions \cite{ga:94} over a huge range by varying $n$.  In contrast to 
amorphous solids, with which ultracold
Rydberg gases share some similarities, the atoms
are practically stationary on the timescale of electronic
dynamics because of their low thermal kinetic energy \cite{anve+:98,moco+:98}.

A striking consequence of the strong Rydberg-Rydberg interaction is
the so-called ''dipole blockade'', i.e.,  a suppression of Rydberg excitations due
to an induced dipole coupling of the Rydberg atoms to their
environment.  This phenomenon was first considered theoretically in
proposals to build fast quantum logic gates \cite{lufl+:01}, to improve the resolution of atomic clocks \cite{bomo:02} and to create  single-atom and single-photon sources \cite{sawa:02}.  It was
experimentally verified for second-order dipole-dipole (or van der Waals) 
coupling between the Rydberg atoms \cite{tofa+:04,sire+:04} by measuring the
density of the Rydberg atoms as a function of increasing laser
intensity, atomic density or principal quantum number, i.e., as a
function of increasing interaction strength.  By applying and varying an 
external electric field the blockade effect was also demonstrated for a direct 
(i.e. first order) dipole-dipole interaction of the Rydberg atoms and it was shown 
that the suppression of excitations is particularly 
pronounced at the so called F\"orster resonances \cite{vovi+:06}. Furthermore, it was
shown that the blockade effect also leads to a quenching of the
probability distribution for the excitation to the Rydberg state
\cite{lire+:05,atpo+:06,hero:06}.

The theoretical description of this laser-driven, interacting
many-particle system is challenging.  In
\cite{tofa+:04} a mean field approach was used  and  the Bloch
equations for a single Rydberg atom in a sphere were solved. Within the 
sphere, embedded in a constant background density of Rydberg atoms, no further 
excitations were allowed.  With the help of a fit parameter the
experimental results of \cite{tofa+:04} could be reproduced.

The system was also investigated by solving the many-particle
Schr\"odinger equation numerically \cite{rohe:05}.  There, intelligent
use was made of the fact that the blockade itself reduces the number
of atoms which can be excited which allows a substantial reduction in
the number of states that had to be considered for the calculations.
Yet, the number of atoms that could be simulated was still so small
that appropriate boundary conditions had to be used to establish
contact with the experiments.  However, experiments using a two-step
(three-level) excitation scheme could not be described since
important effects, such as radiative decay, were not included.

Here, we focus in particular on the two-step excitation scheme, used
in the experiments \cite{sire+:04,lire+:05}, where the intermediate
level decays radiatively.  As we will show, this leads to a reduction of the description of the Rydberg
excitation dynamics in a single atom to a rate equation which in
turn enables us to formulate a quasi-classical approach taking fully
into account all atoms in the excitation volume and all interactions
of the Rydberg atoms.

Experimentally, a gas of atoms is prepared in a magneto-optical
trap (MOT) with peak densities up to $10^{11}$ cm$^{-3}$ at
temperatures of about $100 \mu$K. Under these conditions the gas is
far from the quantum degenerate regime and can be viewed as a
classical ideal gas.  Furthermore, the laser intensities used in
\cite{sire+:04} and \cite{lire+:05} are relatively low, so that
coherent coupling of the atoms by the laser field, e.g., through
stimulated emission and reabsorption of photons, is negligible.
However, the interaction of the individual atoms with the laser fields
has to be treated quantum mechanically.

Our approach is based on the observation that, under the conditions of the
experiments \cite{sire+:04} and \cite{lire+:05}, the description of the
single-atom excitation dynamics can be reduced substantially to a single rate
equation using an adiabatic approximation for the coherences.  Despite the
approximations made, the rate equation accurately describes the population
dynamics of the Rydberg state, including non-trivial effects like the
Autler-Townes splitting of the excitation line. This simplification in the
description of the single-atom dynamics is the key that ultimately allows us to
fully account for the correlated many-particle dynamics with a simple Monte-Carlo
sampling, thereby reducing greatly the complexity of a full quantum treatment.

The paper is organized as follows. In Sec. \ref{model} we present the approach,
 which enables us to describe the dynamics in an ultracold gas of
interacting three-level atoms using a many-body rate equation. Starting from the
full quantum master equation, we justify our approximations first on
the single-atom level (Sec. \ref{nisys}), then for the interacting system (Sec.
\ref{cmpdyn}) and finally describe how the Rydberg-Rydberg interaction is included in
our description (Sec. \ref{ryryint}). For two interacting atoms, we compare the
results of our rate equation with the solution of the quantum master equation
(Sec. \ref{twoatom}). In Sec. \ref{exper} we compare the results of the
simulations for a realistic number of atoms with experimental data and comment on
the possibility to experimentally observe an interaction induced enhancement of
Rydberg excitation (``antiblockade''). Section \ref{final} summarizes 
the results.

Throughout the paper atomic units will be used unless stated otherwise.

\section{Two-step Rydberg excitation in an ultracold gas}\label{model}
\subsection{Dynamics of the non-interacting system}\label{nisys}
\begin{figure}
    \centering
    \includegraphics[width=0.6\columnwidth]{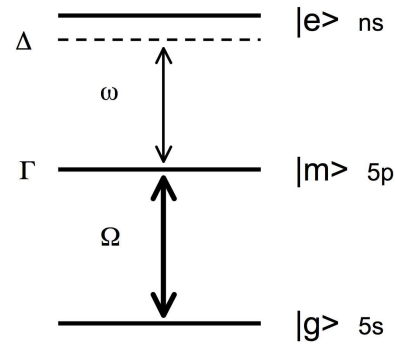}
    \caption{Sketch of the two-step excitation scheme for rubidium.}
    \label{fig:sketch}
\end{figure}

In what follows, we will discuss a two-step cw-excitation scheme for
the Rydberg state (see \fig{fig:sketch}), as typically used in
experiments.  In the first step, the atom is excited from its ground
state $|g\rangle$ to an intermediate level $|m\rangle$ with a
transition strength given by the Rabi frequency $\Omega$.  The photon
for this step is typically provided by the MOT lasers, which are tuned
on resonance with the transition $|g\rangle \to |m\rangle$ during the
time of Rydberg excitation.

In the second step, a separate tunable laser drives the transition
between the intermediate level and the desired Rydberg state
$|e\rangle$ with Rabi frequency $\omega$, where in this step we allow
for a detuning $\Delta$ from resonant excitation.  The decay of the
intermediate level with rate $\Gamma$ has to be taken into account, as
its radiative lifetime is typically shorter than the pulse duration.
On the other hand, the lifetime of the Rydberg state is much longer so
that its decay can be neglected.

The coherent dynamics of $N$ non-interacting three-level atoms coupled
to the two laser fields is governed in
the interaction picture by the Hamiltonian $H_0$,
\begin{equation}
\label{h0}
H_0 = H_{\Delta} + H_{AL} \equiv \sum_i^N h_{\Delta}^{(i)} + \sum_i^N h_{AL}^{(i)} \, ,
\end{equation}
where
\begin{subequations}
\begin{eqnarray}
h_{\Delta}^{(i)} &=& \Delta \, |e_i\rangle\langle e_i| \; ,\\
h_{AL}^{(i)} &=& \frac{\Omega}{2} \left(|m_i\rangle\langle g_i| +
|g_i\rangle\langle m_i|\right) \nonumber \\
 & & {}+ \frac{\omega}{2} \left(|e_i\rangle\langle m_i| +
|m_i\rangle\langle e_i|\right)
\end{eqnarray}
\end{subequations}
describe the interaction of the levels of atom $i$ with the laser beams.

The time evolution of the system including the decay of the
intermediate level is then given by a quantum master equation for the
$N$-particle density matrix $\hat{\rho}^{(N)} \equiv
\hat{\boldsymbol{\rho}}$,
\begin{equation}
\label{ME}
\frac{{\rm d}}{{\rm d}t} \hat{\boldsymbol{\rho}} = -i \left[H_0,\hat{\boldsymbol{\rho}}\right] +
\mathcal{L}\left[\hat{\boldsymbol{\rho}}\right] \; ,
\end{equation}
where the spontaneous decay of level $|m\rangle$ is included via the
Lindblad operator $\mathcal{L}$.  In general, the rate of spontaneous
decay of an atom is influenced by the presence of other atoms through
a coupling mediated by the radiation field, which can account for
collective effects like superradiance.  The strength of this coupling
is determined by the dimensionless quantity $x_{ij} \equiv 2\pi
|\mathbf{r}_i - \mathbf{r_j}|/\lambda$, which measures the atom-atom
distance in units of the wavelength $\lambda$ of the $|g\rangle \to
|m\rangle$ transition.  For $x_{ij} \ll 1$ the spontaneous decay of an
atom is strongly affected by its neighbors, while for $x_{ij}\gg 1$
the atoms radiate independently.
In typical experiments with ultracold gases, the mean atomic distance
between atoms is $a\sim 5 \mu$m.  For the 5$s$ $\to$ 5$p$ transition of Rb
this corresponds to $x_{ij} \sim 40$.  Hence, the collective decay is
negligible and the Lindblad operator can be written as a sum of single-atom operators,
\begin{equation}
\label{Lindblad_single}
\mathcal{L} = \Gamma \sum_i^N \left( L_i \hat{\boldsymbol{\rho}}
L^{\dagger}_i - \frac{1}{2} L^{\dagger}_i L_i \hat{\boldsymbol{\rho}} -
\frac{1}{2} \hat{\boldsymbol{\rho}} L^{\dagger}_i L_i\right)
\end{equation}
with
\begin{equation}
L_i = |g_i\rangle\langle m_i| \quad \text{and} \quad
L^{\dagger}_i=|m_i\rangle\langle g_i| \; .
\end{equation}
Hence, the dynamics of the atoms is completely decoupled and the $N$-atom density
matrix factorizes, $\hat{\boldsymbol{\rho}} = \hat{\rho}^{(1)}_1 \otimes\, \dots
\, \otimes \hat{\rho}^{(1)}_N$. The time evolution of a non-interacting gas of
three-level atoms is therefore completely determined by the master equation for the
single-atom density matrix $\hat{\rho}^{(1)}_k \equiv \hat{\rho}$, i.e., the
optical Bloch equations (OBE) for a three-level atom,
\begin{subequations}
\label{obe}
\begin{eqnarray}
\dot{\rho}_{\text{gg}} &=& 
i\frac{\Omega}{2}\left( \rho_{\text{gm}} - \rho_{\text{mg}}\right)
+\Gamma \rho_{\text{mm}} \\[1ex]
\dot{\rho}_{\text{mm}} &=&
-i\frac{\Omega}{2}\left( \rho_{\text{gm}} -
\rho_{\text{mg}}\right) \nonumber \\
 & & +i\frac{\omega}{2}\left( \rho_{\text{me}} - \rho_{\text{em}}\right)
-\Gamma \rho_{\text{mm}} \\[1ex]
\dot{\rho}_{\text{ee}} &=& 
-i\frac{\omega}{2}\left( \rho_{\text{me}} - \rho_{\text{em}}\right)
\\[1ex]
\dot{\rho}_{\text{gm}} &=& 
-i \frac{\Omega}{2} \left( \rho_{\text{mm}} -
\rho_{\text{gg}}\right) 
+i \frac{\omega}{2}  \rho_{\text{ge}}
-\frac{\Gamma}{2} \rho_{\text{gm}} \\[1ex]
\dot{\rho}_{\text{me}} &=& 
-i \Delta \rho_{\text{me}}
-i \frac{\omega}{2} \left( \rho_{\text{ee}} -
\rho_{\text{mm}}\right) \nonumber \\
 & & -i \frac{\Omega}{2}  \rho_{\text{ge}}
-\frac{\Gamma}{2} \rho_{\text{me}}  \\[1ex]
\dot{\rho}_{\text{ge}} &=& 
-i \Delta  \rho_{\text{ge}}
- i \frac{\Omega}{2}  \rho_{\text{me}}
+ i \frac{\omega}{2}  \rho_{\text{gm}}
\\[1.5ex]
\rho_{\beta\alpha} &=& \left( \rho_{\alpha\beta} \right)^{\star} \qquad \mbox{for} \quad
\alpha \ne \beta \, .
\end{eqnarray}
\end{subequations}
As usual, the level populations are described by the diagonal elements of the
density matrix, whereas the off-diagonal elements, i.e., the coherences, contain the 
information about the transition amplitudes between the levels. 
Conservation of probability leads to the sum rule 
\begin{equation}
\label{sumrule}
\sum_{\alpha} \rho_{\alpha\alpha}=1
\end{equation}
for the populations so that 8 independent variables remain to be solved for.

This single-atom description is too complex to serve as the
basis of a tractable description for the many-particle system.
Fortunately, under the set of  relevant experimental parameters \eq{obe} 
simplifies substantially.  In the experiments \cite{sire+:04,lire+:05},
the upper transition is much more weakly driven than the
lower one ($\omega \ll \Omega$) due to the different transition
dipole moment of the respective excitation.  This defines two well
separated time scales, such that the Rydberg transition
$|m\rangle \to |e\rangle$ is slow compared to the pump transition
$|g\rangle \to |m\rangle$.
Thus, the time evolution of the system is governed by the slow Rydberg
transition in the sense that the coherences of the fast pump
transition will adiabatically follow the slow dynamics of the Rydberg
transition.

Furthermore, the decay rate of the intermediate level is much larger
than the Rabi frequency of the upper transition ($\Gamma \gg \omega$)
implying that the populations will evolve only slightly over a time
$\Gamma^{-1}$.  Hence, dephasing of the atomic transition dipole
moments, i.e., damping of the oscillations of coherences, is fast
compared to the dynamics of the Rydberg population.

Under these conditions, the coherences can be expressed as a function
of the populations at each instant of time, i.e., their dynamics can
be eliminated adiabatically \cite{codu+:92} by setting
\begin{equation}
\label{adiab_approx}
\dot{\rho}_{\alpha\beta}=0 \qquad \mbox{for} \quad \alpha \ne \beta \, .
\end{equation}
Solving the algebraic equations arising from (\ref{obe}) and
(\ref{adiab_approx}) for the populations, making use of (\ref{sumrule}) and
inserting into the differential equation for $\rho_{\text{mm}}$ and
$\rho_{\text{ee}}$ one arrives at
\begin{subequations}
\label{prelim_re}
\begin{eqnarray}
\dot{\rho}_{\text{mm}} & = & q_1 \rho_{\text{mm}} + q_2 \rho_{\text{ee}} + q_3 \\
\dot{\rho}_{\text{ee}} & = & q_4 \rho_{\text{mm}} + q_5 \rho_{\text{ee}} + q_6
\, ,
\end {eqnarray}
\end{subequations}
where the coefficients $q_k=q_k(\Omega ,\omega ,\Gamma ,\Delta)$ are some functions of the 
parameters describing the excitation dynamics of the three-level system.

To simplify further, we note that within the adiabatic approximation
(\ref{adiab_approx}) the dynamics of the population difference $\rho_{\text{mm}}
- \rho_{\text{gg}}$ can be neglected for times $t > 1/2\Gamma$. This can be
  verified by direct integration of $\dot{\rho}_{\text{mm}} -
  \dot{\rho}_{\text{gg}}$ from the OBE, which shows that the dynamics of the
  population difference is proportional to $1-\exp(-2\Gamma t)$ and thus reaches
  its saturation limit at a timescale on the order of $1/2\Gamma$. Using the sum
  rule (\ref{sumrule}) this leads to the relation
\begin{equation}
 2\dot{\rho}_{\text{mm}} + \dot{\rho}_{\text{ee}}=0 \, ,
\end{equation}
which can be used to eliminate the population of the intermediate level occurring
in (\ref{prelim_re}). 
Finally, one arrives at a single differential equation for
$\rho_{\text{ee}}$
\begin{equation}
\label{diff_eq}
\dot{\rho}_{\text{ee}}=-\frac{\gamma_{\uparrow}}{\rho_{\text{ee}}^{\infty}} \rho_{\text{ee}}
+ \gamma_{\uparrow} \, ,
\end{equation}
which can readily be solved to give
\begin{equation}\label{sol}
\rho_{\text{ee}}(t) =\rho_{\text{ee}}^{\infty} \left( 1 - \exp\left[-
\frac{\gamma_{\uparrow}}{\rho_{\text{ee}}^{\infty}} t \right] \right) \, ,
\end{equation}
where $\rho_{\text{ee}}^{\infty}=\rho_{\text{ee}}^{\infty}(\Omega ,\omega
,\Gamma ,\Delta)$ denotes the steady-state occupation of level $|e\rangle$ and
$\gamma_{\uparrow}=\gamma_{\uparrow}(\Omega ,\omega ,\Gamma ,\Delta)$ is the rate for populating the
Rydberg level for short times.  The expressions for $\gamma_{\uparrow}$ and
$\rho_{\text{ee}}^{\infty}$ are given in the Appendix. Here we 
note only 
that in the limit $\Omega \gg \Gamma \gg \omega $ they reduce to
\begin{equation}
\gamma_{\uparrow}=\frac{\Gamma\omega^{2}/\Omega^{2}}
{2(1-4\Delta^{2}/\Omega^{2})^{2}}\quad , \quad
\rho_{\text{ee}}^{\infty}=\frac{1}{1+8\Delta^{2}/\Omega^{2}}\, ,
\end{equation}
which shows that  the resonant excitation rate is proportional to
$(\omega/\Omega)^2$.

Introducing an effective ground state $\rho_{\text{gg}}^{\text{eff}} = 1 -
\rho_{\text{ee}}$, one can write (\ref{diff_eq}) in the form of a rate equation (RE) for
an effective two-level atom
\begin{equation}
\label{re}
\dot{\rho}_{\text{ee}}(t) = \gamma_{\uparrow}\, \rho_{\text{gg}}^{\text{eff}} -
\gamma_{\downarrow}\, \rho_{\text{ee}} \, ,
\end{equation}
with de-excitation rate 
\begin{equation}
\gamma_{\downarrow} = \gamma_{\uparrow}\, \left(
\frac{1-\rho_{\text{ee}}^{\infty}}{\rho_{\text{ee}}^{\infty}} \right) \, .
\end{equation}

\begin{figure}
   \centering
   \includegraphics[width=0.7\columnwidth]{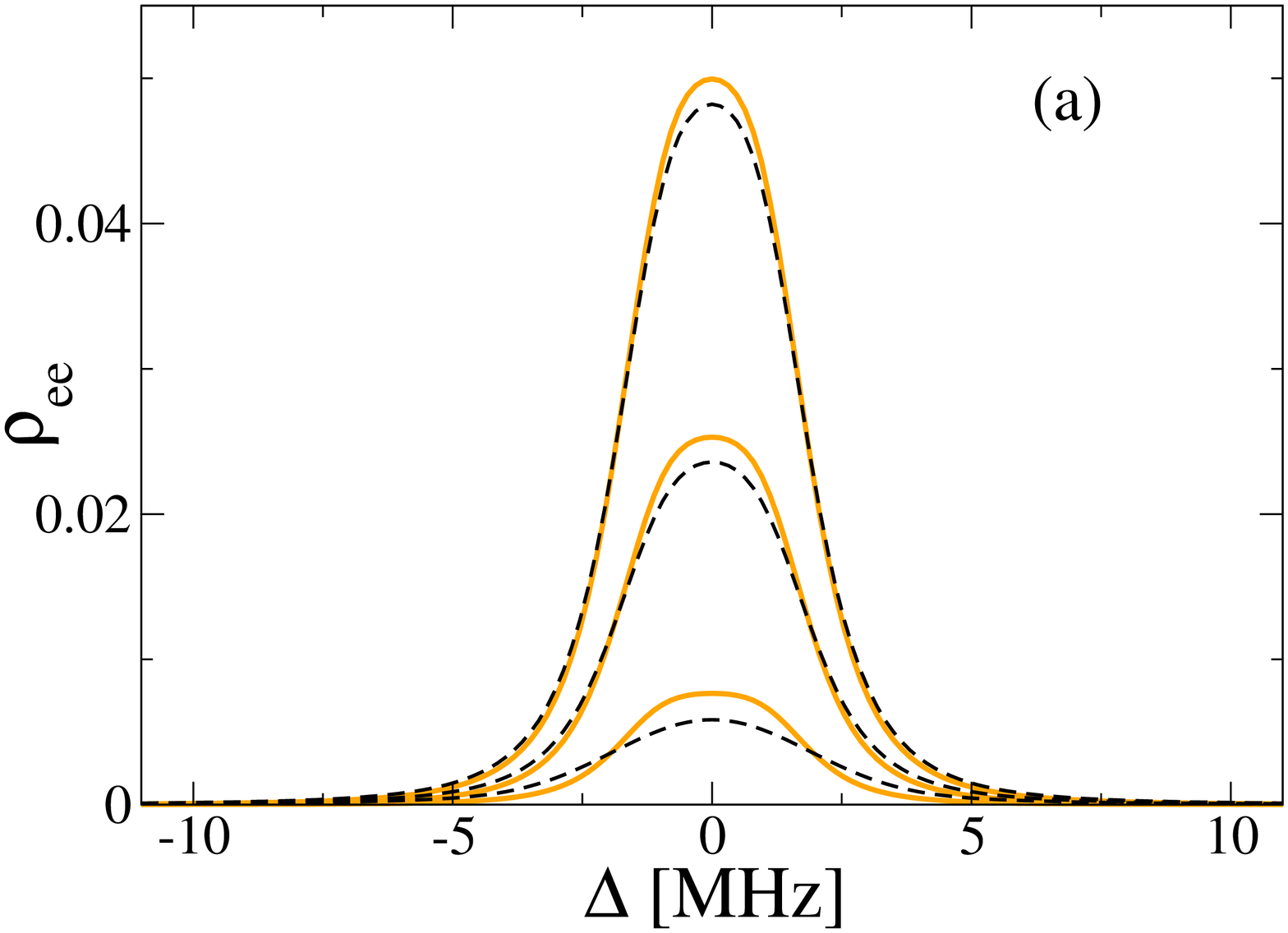}
   \includegraphics[width=0.7\columnwidth]{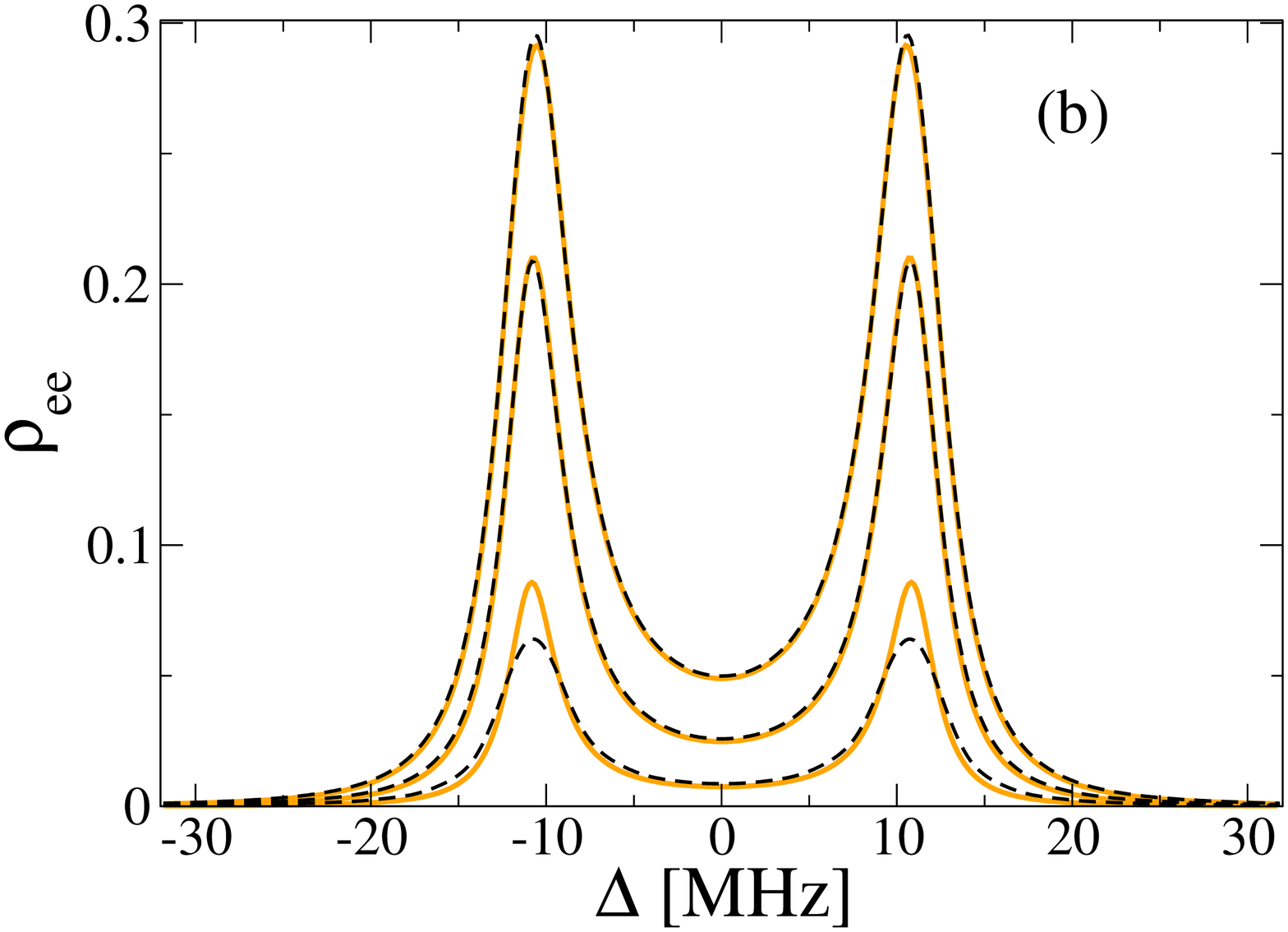}
   \caption{Population of the Rydberg level for the three-level system of Fig.
   \ref{fig:sketch} according to the RE \eq{re} (solid lines) and OBE 
   \eq{obe} (dashed lines) for
   different pulse lengths: 0.3 $\mu$s (lowest pair of curves), 1.0 $\mu$s (middle
   pair) and 2.0 $\mu$s. The parameters in MHz are: (a) 
   $(\Omega,\omega,\Gamma) = (4, 0.2, 6)$ in (a) and $(22.1, 0.8, 
   6)$ in (b).} 
   \label{fig:populations}
\end{figure}

A comparison of the solutions of the OBE (\ref{obe}) and the RE
(\ref{re}) for the Rydberg populations as function of the detuning
$\Delta$ is shown in \fig{fig:populations} for different pulse
lengths.  The parameters correspond to those of the experiments
\cite{lire+:05} and \cite{sire+:04}.  The agreement of the solutions
is generally good and becomes even better for longer pulses.  For the
parameters of experiment \cite{lire+:05} the convergence of the RE
solution to that of the OBE in the region around $\Delta=0$ is slower
as a function of pulse length.  This is due to $\Omega < \Gamma$ which
indicates that it is not fully justified to neglect the nonlinear
short-time population dynamics.

The RE reproduces the Autler-Townes splitting of the intermediate level $| m \rangle$ manifest in a splitting of the Rydberg line,
proportional to $\Omega$ for short times. The splitting is transient, 
as the steady state with its single central peak  is approached for long 
times when the Rydberg population reaches the saturation limit. A detailed
analysis of the peak structure of the Rydberg populations in this system,
especially the occurrence of the Autler-Townes splitting and its impact on the
excitation blockade, has been given in \cite{atpo+:07a}. 

For future reference, we will cast the single-atom RE (\ref{re}) into a form
which will be used for the simulation of the interacting many-particle system.
To this end, we denote the state of the atom
by $\sigma$, where $\sigma=1$ if
the atom is in the Rydberg state, and $\sigma=0$ otherwise. 
Furthermore, we define the rate of  change for the state $\sigma$,
\begin{equation}
\gamma(\Delta,\sigma) \equiv (1-\sigma)\: \gamma_{\uparrow}(\Delta) + \sigma \:
\gamma_{\downarrow}(\Delta) \, , 
\end{equation}
which describes excitation of the atom if it is in the (effective)
ground state $(\sigma=0)$ and  de-excitation if it is in the
excited state ($\sigma=1$).  Using these definitions, we can combine
(\ref{re}) which determines $\rho_{\text{ee}}(t)$ and the corresponding equation for
${\rho}_{\text{gg}}^{\text{eff}}(t)$ in the form of an evolution
equation for the single-atom state distribution function $p(\sigma)$,
\begin{equation}
\label{single_sdf}
\frac{dp(\sigma)}{dt} = \sum_{\sigma^{\prime}} \, T(\sigma,\sigma^{\prime}) \, p(\sigma^{\prime})
\end{equation}
with $p(0) = 1 -
\rho_{\text{ee}}$, $p(1)=\rho_{\text{ee}}$ and the transition rate matrix
\begin{equation}
\label{TMsingle}
T(\sigma,\sigma^{\prime}) =  -\gamma (\Delta,\sigma)\,\delta_{\sigma,\sigma^{\prime}}
+ \gamma (\Delta,1-\sigma)\,\delta_{1-\sigma,\sigma^{\prime}} \, .
\end{equation}
The first term of (\ref{TMsingle}) describes the transition $\sigma\to 1-\sigma$, through which the system can leave the state $\sigma$, while the opposite process  ($1 -\sigma \to \sigma$), which brings the system into the state $\sigma$, is described by the second term.

Proceeding to the case of $N$ non-interacting atoms, we 
define the many-particle state
$\boldsymbol{\sigma}$ as the configuration containing all single-atom states $\sigma_i$,
i.e., $\boldsymbol{\sigma} \equiv \{ \sigma_1, \dots , \sigma_i, \dots \sigma_N
\}$ and $\boldsymbol{\sigma}_i$ as
the many-body configuration which is identical to $\boldsymbol{\sigma}$ except
for the state of atom $i$, i.e., $\boldsymbol{\sigma}_i \equiv \{\sigma_1,
\dots, 1-\sigma_i, \dots, \sigma_N \}$. 
If we finally use the notation $\boldsymbol{\gamma} (\Delta,\boldsymbol{\sigma}) \equiv \sum_i \gamma
 (\Delta, \sigma_i)$ and $\delta_{\boldsymbol{\sigma},\boldsymbol{\sigma}^{\prime}} \equiv \delta_{\sigma_1,\sigma_1^{\prime}} \cdots \delta_{\sigma_N,\sigma_N^{\prime}}$,
the matrix of the transition rates generalizes to
\begin{equation}
\label{m_sdf}
T(\boldsymbol{\sigma},\boldsymbol{\sigma}^{\prime}) = 
 -\boldsymbol{\gamma} (\Delta,\boldsymbol{\sigma})\,\delta_{\boldsymbol{\sigma},\boldsymbol{\sigma}^{\prime}}
 + \sum_i \gamma (\Delta,1-\sigma_i)\,\delta_{\boldsymbol{\sigma}_i,\boldsymbol{\sigma}^{\prime}} \, ,
\end{equation}
and the evolution equation for the many-body state distribution function $P(\boldsymbol{\sigma})$ can be written in a closed form as
\begin{equation}
\label{many_sdf}
\frac{dP(\boldsymbol{\sigma})}{dt}  = \sum_{\boldsymbol{\sigma}^{\prime}} \, T(\boldsymbol{\sigma},\boldsymbol{\sigma}^{\prime}) P(\boldsymbol{\sigma}^{\prime}) \; .
\end{equation}
For non-interacting particles the rate $\gamma$  depends (besides on the laser detuning) only on the
state of particle $i$, i.e., on $\sigma_i$. However, this is no longer true in the interacting case
and $\gamma$ will depend on the entire many-body configuration.

\subsection{Correlated many-particle dynamics}\label{cmpdyn}

In order to study the correlated dynamics of the interacting many-particle system, we have to add the
Hamiltonian describing the Rydberg-Rydberg interaction
\begin{equation}
H_{RR} = \frac{1}{2} \sum_{i,j \, (i\ne j)} U_{ij}\, |e_i, e_j \rangle\langle
e_i,e_j |
\end{equation}
to $H_0$ (c.f. eq.\ (\ref{h0})), where $U_{ij}$ is the interaction energy of a pair
of Rydberg atoms at a distance $r_{ij} \equiv |\mathbf{r}_i-\mathbf{r}_j|$.
The quantum master equation (\ref{ME}) then reads
\begin{equation}\label{fullME}
\frac{{\rm d}}{{\rm d}t} \hat{\boldsymbol{\rho}} = -i \left[H_0 + H_{RR},
\hat{\boldsymbol{\rho}}\right] + \mathcal{L}\left[\hat{\boldsymbol{\rho}}\right] \; ,
\end{equation}
with the Lindblad operator given in (\ref{Lindblad_single}).

To see which terms of the master equation are affected by the inclusion of the
Rydberg-Rydberg interaction we consider the commutator $[H_{\Delta} + H_{RR},
\boldsymbol{\rho}]$ in the many-body basis $|\boldsymbol{\alpha}\rangle \equiv | \alpha_1, \dots
\alpha_N \rangle = |\alpha_1\rangle \cdots |\alpha_N\rangle$, where
$|\alpha_i\rangle$ denotes the state of atom $i$,
\begin{widetext}
\begin{equation}
\left( [H_{\Delta} + H_{RR}, \boldsymbol{\rho}] \right)_{\boldsymbol{\alpha} 
\boldsymbol{\beta}} 
  =  \sum_i \Bigg[ \Big( \Delta + \sum_{j \, (i\ne j)}
\frac{U_{ij}}{2} \delta_{\alpha_j ,e} \Big) \delta_{\alpha_i, e} 
  - \Big( \Delta + \sum_{j \, (i\ne j)}
\frac{U_{ij}}{2} \delta_{\beta_j ,e} \Big) \delta_{\beta_i,
e} \Bigg] \boldsymbol{\rho}_{\boldsymbol{\alpha} \boldsymbol{\beta}}
\, ,
\end{equation}
\end{widetext}
and rewrite it (using the conservation of probabilities for each atom,
i.e., $1=\delta_{\mu_k,g}+\delta_{\mu_k,m}+\delta_{\mu_k,e} \equiv 
\delta_{\mu_k,\tilde{g}}+\delta_{\mu_k,e}$, and the symmetry of the
Rydberg-Rydberg interaction, $U_{ij}=U_{ji}$) as
\begin{widetext}
\begin{eqnarray}\label{commutator}
\left( [H_{\Delta} + H_{RR}, \boldsymbol{\rho}] \right)_{\boldsymbol{\alpha} 
\boldsymbol{\beta}} &=& 
\sum_i \left(
\delta_{\alpha_i,e} \, \delta_{\beta_i,\tilde{g}} -
\delta_{\alpha_i,\tilde{g}} \, \delta_{\beta_i,e}
\right)
\left[
\Delta + \sum_{j\ne i} U_{ij}\, \delta_{\alpha_j,e} \, \delta_{\beta_j,e}
\right]\boldsymbol{\rho}_{\boldsymbol{\alpha} \boldsymbol{\beta}} 
\nonumber \\
& & +
\sum_{i,j\,(i\ne j)} \frac{U_{ij}}{2}
\left(
\delta_{\alpha_i,e}\, \delta_{\alpha_j,e}\, \delta_{\beta_i,\tilde{g}}\,
\delta_{\beta_j,\tilde{g}} -
\delta_{\alpha_i,\tilde{g}}\, \delta_{\alpha_j,\tilde{g}}\,
\delta_{\beta_i,e}\, \delta_{\beta_j,e}
\right)\boldsymbol{\rho}_{\boldsymbol{\alpha} \boldsymbol{\beta}}
\; .
\end{eqnarray}
\end{widetext}

In the first term of (\ref{commutator}) the Rydberg-Rydberg interaction shows up
as an additional (local) detuning of an atom at $\mathbf{r}_i$, whenever the
atom at $\mathbf{r}_j$ is in the Rydberg state (i.e., if $\alpha_j =
\beta_j=e$). In particular, no additional coherences are
generated by the Rydberg-Rydberg interaction and, therefore, this term does not
change the structure of the master equation as compared to the non-interacting
case.

The second term describes direct transitions between states where atoms $i$ and
$j$ are \emph{not} in the Rydberg state and the state where the atoms form a
Rydberg pair. These transitions require the simultaneous absorption or emission
of at least two photons and are thus higher order processes. The dynamics of
these multi-photon processes is very slow compared to all other transitions in
the system, therefore it can be neglected (see also the discussion in sections
\ref{twoatom} and \ref{dblock}), i.e., the commutator (\ref{commutator}) can be
approximated as
\begin{widetext}
\begin{eqnarray} \label{det_operator}
\left( [H_{\Delta} + H_{RR}, \boldsymbol{\rho}] \right)_{\boldsymbol{\alpha} 
\boldsymbol{\beta}} 
&\approx& \sum_i \left(
\delta_{\alpha_i,e} \, \delta_{\beta_i,\tilde{g}} -
\delta_{\alpha_i,\tilde{g}} \, \delta_{\beta_i,e}
\right)
\left[
\Delta + \sum_{j\ne i} U_{ij}\, \delta_{\alpha_j,e} \, \delta_{\beta_j,e}
\right]\boldsymbol{\rho}_{\boldsymbol{\alpha} \boldsymbol{\beta}}
\; .
\end{eqnarray}
\end{widetext}
Thus, within this approximation, we recover the simple picture which is commonly
used for the explanation of the dipole blockade effect, namely that a highly
excited atom shifts the Rydberg levels of nearby atoms out of resonance with the
excitation laser.

By neglecting multi-photon transitions, the structure of the master equation is
not changed compared to the non-interacting system and we can perform the
adiabatic approximation discussed above. Identifying finally $\delta_{\alpha_j,e} \,
\delta_{\beta_j,e}$ with $\sigma_j$, it is straightforward to generalize Eq.
(\ref{m_sdf}) to the interacting case,
\begin{equation}
\label{TMmany}
T(\boldsymbol{\sigma},\boldsymbol{\sigma}^{\prime}) = 
 -\boldsymbol{\gamma} (\Delta,\boldsymbol{\sigma})\,\delta_{\boldsymbol{\sigma},\boldsymbol{\sigma}^{\prime}}
 + \sum_i \gamma (\Delta_i,1-\sigma_i)\,\delta_{\boldsymbol{\sigma}_i,\boldsymbol{\sigma}^{\prime}} \, ,
\end{equation}
where now $\boldsymbol{\gamma}(\Delta,\boldsymbol{\sigma}) = \sum_i \gamma (\Delta,\delta_i,\sigma_i)$ and all atoms are coupled by the energetic shift caused by the Rydberg-Rydberg interaction
\begin{equation}
\Delta_i = \Delta + \delta_i \equiv \Delta + \sum_{j\ne i} \sigma_j\: U_{ij} \, ,
\end{equation}
so that in the interacting case the rate for a state change $\gamma (\Delta,\delta_i,\sigma_i)$ for the atom $i$ depends on the entire many-body configuration through the local detuning $\delta_i$.

The above approximations simplify the description of the correlated many-particle dynamics to a high degree, since a particular many-particle configuration $\boldsymbol{\sigma}$ is directly coupled to ''only'' $N$  configurations $\boldsymbol{\sigma}^{\prime}$ by the transition rate matrix $T(\boldsymbol{\sigma},\boldsymbol{\sigma}^{\prime})$, which has to be compared to the available number of $2^N$ many-particle states.
To explicitly show this simplification, we insert (\ref{TMmany}) into the evolution equation (\ref{many_sdf}) of the state distribution function, perform the sum over $\boldsymbol{\sigma}^{\prime}$ and finally arrive at
\begin{equation}
\label{many_RE}
\frac{dP(\boldsymbol{\sigma})}{dt}   =  - \sum_i^N \gamma(\Delta,\delta_i,\sigma_i) \:
P(\boldsymbol{\sigma}) + \sum_i^N \gamma (\Delta,\delta_i,1-\sigma_i) \:
P(\boldsymbol{\sigma}_i) \; .
\end{equation}
Knowing $U_{ij}$, Eq. (\ref{many_RE}) can be solved with standard Monte-Carlo
sampling techniques, allowing us to treat systems up to several $10^5$ atoms.

We emphasize that the description presented above is \emph{not} restricted to the three-level scheme considered in this work. It can, e.g., also be applied for a direct excitation of the Rydberg state from the ground state (two-level scheme) provided that the atomic coherences are damped out fast enough to not  significantly affect the population dynamics of the Rydberg state (e.g., if the bandwidth of the excitation laser is larger than the Rabi frequency of the transition). For a single-step excitation scheme the (de)-excitation rates are given by 
\[
\gamma_{\uparrow} = \gamma_{\downarrow} = \frac{2\, \Gamma\, \Omega^2}{\Gamma^2+4\,\Delta^2} \, ;
\]
where $\Gamma$ is the measured width of the excitation line.

\subsection{Determination of the Rydberg-Rydberg interaction}
\label{ryryint}
An accurate determination of the interaction potential $U_{ij}$ is challenging
due to the mixing of a large number of electronically excited molecular
potential curves. Results from a perturbative treatment exist  for the $r_{ij}
\to\infty$ asymptote of the alkali-metal atoms \cite{sist+:05} and for the level shifts of Rb \cite{recu+:07} as well as calculations for Cs based on the 
diagonalization of the interaction Hamiltonian of two highly excited atoms using
a large number ($\sim 5000$) of pair states as basis \cite{sccr+:06}. In the
latter spirit, a simple picture was formulated in \cite{lita+:05} for Rb that
allows for an intuitive understanding of the basic dependence of $U_{ij}$ on
$r_{ij}$ and on the principal quantum number $n$ of the Rydberg state.

Following \cite{lita+:05}, 
a pair of Rydberg atoms in states $|a\rangle$ and $|b\rangle$ at distance
$r_{ij}$ experiences a shift $U_{ij}$ of its electronic energy due to an induced
dipole coupling $V_{ij} = \mu_{aa'}\mu_{bb'}/r_{ij}^{3}$ to an energetically
close pair  of states $|a'\rangle$ and $|b'\rangle$. The shift is given by the
eigenvalues
\begin{equation}
\label{pair_interaction}
U_{ij} = \frac{1}{2} 
\left(\delta_0 \pm \sqrt{\delta_0^{2}+4V_{ij}^{2}}\right)
\end{equation}
of the two-state Hamiltonian matrix
\[
\mathcal{H}=
\left(
\begin{array}{cc}
 0 & V_{ij} \\
 V_{ij} & \delta_0
\end{array}
\right) \, ,
\]
where $\delta_0$ is the asymptotic ($r_{ij}\to\infty$) difference between the
energies of the two pairs.

For a pair $|ns,ns\rangle$ of two atoms in the $ns$ state, the relevant dipole
coupling is to the energetically close pair $|(n-1)p_{3/2},\, np_{3/2}\rangle$.
For an arbitrary but fixed quantum number $n_{0}$ we may define $\mu^{2}(n_{0})\equiv
\mu_{n_{0}s(n_{0}-1)p}\mu_{n_{0}sn_{0}p}$.
The interaction strength for other Rydberg levels $n$ then follows from 
the scaling \cite{ga:94}

\begin{subequations}
\label{n-scale}
\begin{eqnarray}
\mu^{2}(n)&=&\mu^{2}(n_{0})
\left(\frac{n^{*}}{n^{*}_{0}}\right)^{4}\\
\delta_0(n) &=& 
\delta_0(n_{0})\left(\frac{n^{*}_{0}}{n^{*}}\right)^{3}\,,
\end{eqnarray}
\end{subequations}
where $n^{*}=n-\eta$ includes the appropriate quantum defect $\eta$ (for the
$ns$ states of Rb $\eta = 3.13$).  For $r_{ij}\to\infty$ one recovers the
familiar van der Waals $r^{-6}$-dependence and the dominant $n^{11}$ scaling for
the pair interaction $U_{ij}$. For Rb we will use in the following the values
$\mu^{2}(n_{0})=843800\,$a.u.\ and
$\delta_0(n_0)= -0.0378\,$a.u.\ for $n_{0}=48$ from \cite{lita+:05}.

\section{An accurate treatment of two interacting atoms}
\label{twoatom}
As a test for our rate equation approach in the case of interacting 
atoms, we have numerically solved the full quantum master equation
(\ref{fullME}) and the rate equation (\ref{many_RE}) for two interacting atoms
separated by an interatomic distance $r$.
\begin{figure}
\centering
\includegraphics[width=\columnwidth]{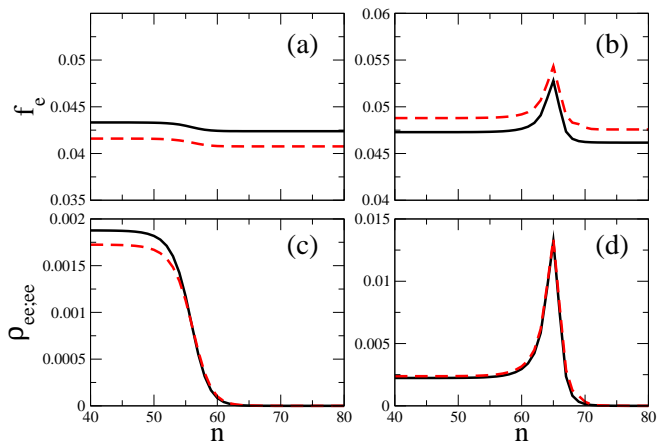}
\caption{Comparison of the solutions of the master equation (\ref{fullME})
(dashed lines) and the rate equation (\ref{many_RE}) (solid lines) for two
interacting atoms at distance $r=5\,\mu$m. Upper graphs (a,b) show the fraction
of excited atoms $f_e$, lower graphs (c,d) the probability $\rho_{ee;ee}$ that
both atoms are in the Rydberg state as function of the principal quantum number
$n$ for a pulse length $\tau=2\,\mu$s. The parameters of (a,c) and (b,d) are
those of Figs.\ \ref{fig:populations}a and  \ref{fig:populations}b, respectively.}
\label{fig:comp}
\end{figure}
The quantity directly accessible in the experiments is the fraction of
excited atoms $f_e$.  It is shown in \fig{fig:comp}a and \fig{fig:comp}b
as a function of the principal quantum number $n$ for excitation
parameters used in the experiments \cite{lire+:05} and
\cite{sire+:04}, respectively.  The overall agreement between the
exact result and our approximation is very good and the discrepancy of
only a few percent between the solutions is comparable to that of the
single-atom calculations (c.f. \fig{fig:populations}; note the
different scaling of the ordinate) and practically independent of the
interaction strength.  This indicates that most of the deviation is a
consequence of the approximations already introduced at the
\emph{single-atom} level.

 For both parameter sets we see a suppression in $f_e$ for large $n$,
 i.e., an excitation blockade.  Additionally, in the case where the
 single-atom excitation spectrum shows a double-peak structure
 [\fig{fig:comp}b], there is an excitation enhancement for a
 certain $n$.  Its actual value depends on the separation $r$ of the
 atoms, so that in a gas this ``antiblockade'' will be smeared out due
 to the wide distribution of mutual atomic distances.  However, for
 atoms regularly arranged in space, i.e., on a lattice where the
 interatomic distances are fixed, the antiblockade should be clearly
 visible \cite{atpo+:07a}.  To verify that the observed
 (anti-)blockade in $f_e$ is really a suppression (enhancement) of
 Rydberg pairs we have plotted the probability $\rho_{ee;ee}$ that
 both atoms are in the Rydberg state.  Indeed, we observe a complete
 suppression of the pair state in the blockade regime
 (\fig{fig:comp}c) and the antiblockade peak
 (\fig{fig:comp}d) as well as a good agreement between the
 solutions of the master and the rate equation in both cases.

\begin{figure*}
\centering
\includegraphics[width=0.9\columnwidth]{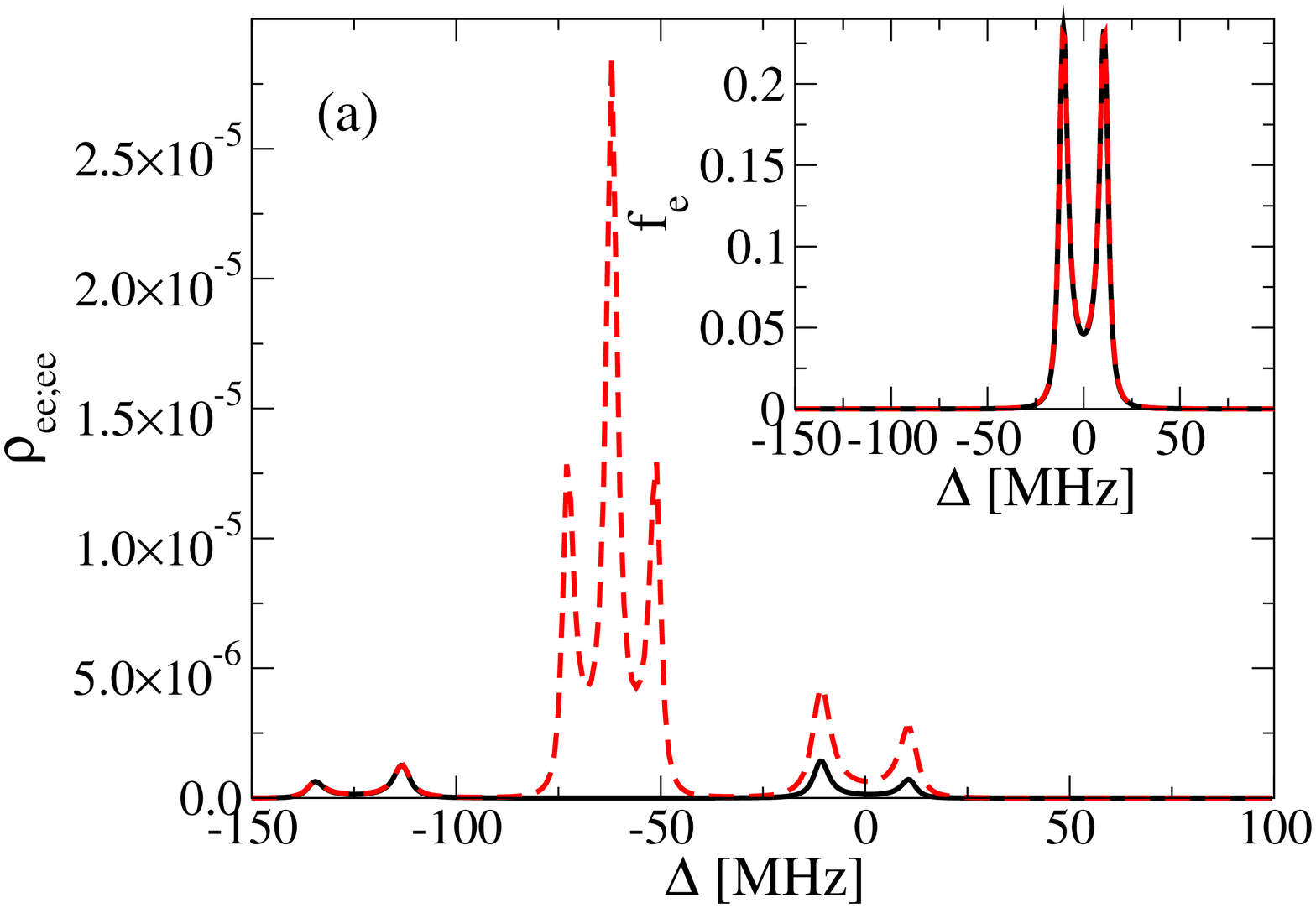}
\includegraphics[width=0.9\columnwidth]{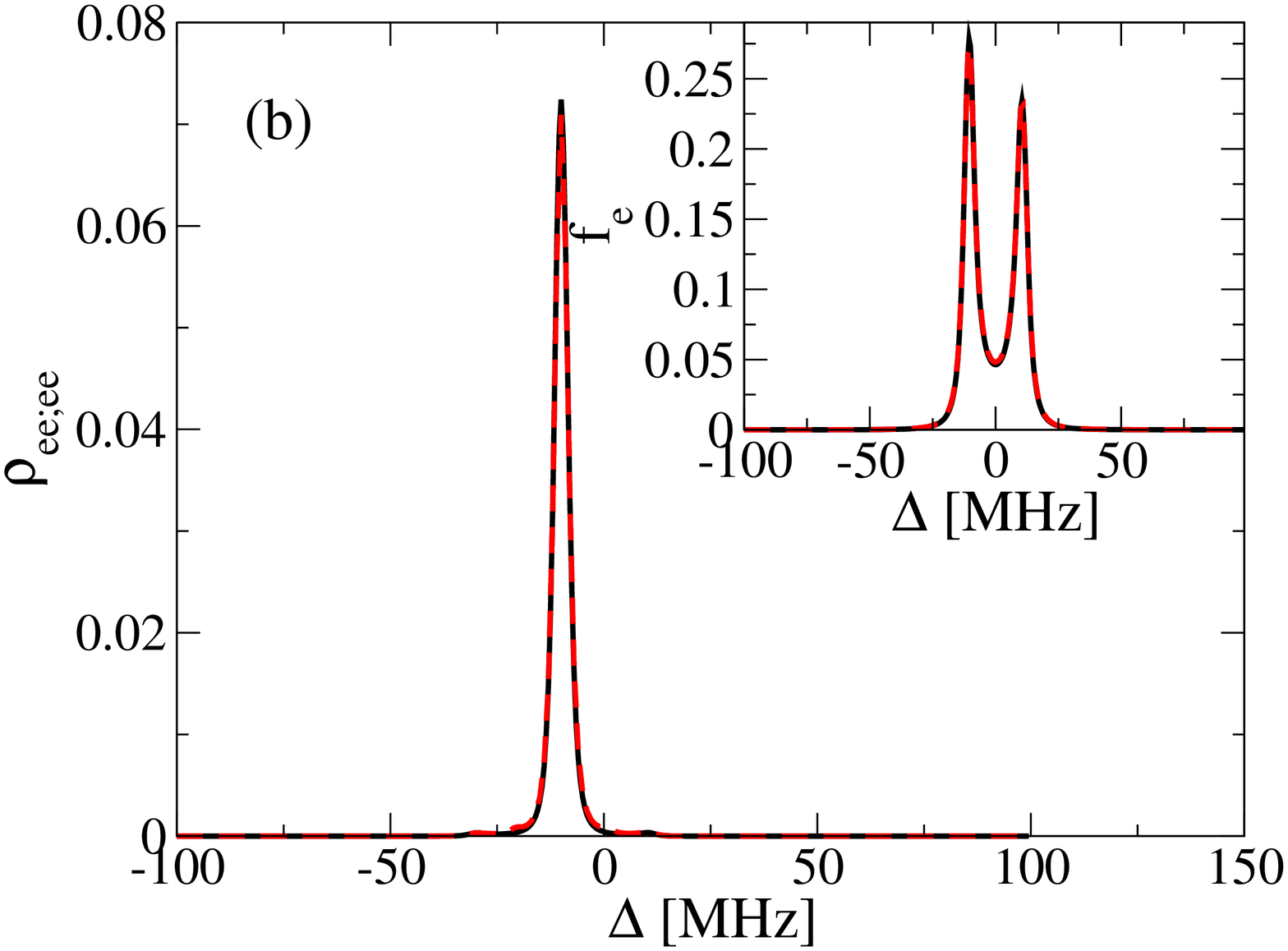}
\caption{Probability $\rho_{ee;ee}$ that both atoms are in the Rydberg
state $82S$ as a function of the laser detuning $\Delta$ after an
excitation time of $\tau=2\,\mu$s at an interatomic distance of
$r=5\,\mu$m (a) and $r=7\,\mu$m (b).  Solid lines are the solutions of
\eq{many_RE}, the dashed lines of \eq{fullME}.  The excitation
parameters are those of \fig{fig:populations}b.  The insets show
the corresponding fraction of excited atoms $f_{e}$.}
\label{fig:twophotonprof}
\end{figure*}

Neglecting two-photon transitions (the second term in \eq{commutator}) is the 
central approximation which
we make in the description of the dynamics of the \emph{interacting}
system.  In fact, these processes can be dominant, if the
two-photon detuning vanishes far away from resonance, i.e., if
$\Delta_{2ph}\equiv 2\Delta + U(r) =0$ for $|\Delta|\gg 0,|U(r)| \gg
0$.  This is clearly seen in \fig{fig:twophotonprof}a, where
$\rho_{ee;ee}$ is shown as a function of the laser detuning $\Delta$
for two atoms separated by $r=5\,\mu$m.  The solution of the master
equation exhibits a triple peak structure with the central peak 
located at
$\Delta=-U(r)/2$ (c.f.\ \eq{commutator}), which is not present
in the solution of the rate equation.  However, the probability for
this two-photon transition is too small to be visible in
the signal of the total probability $f_e$ that the atoms are in the
Rydberg state (see inset).

Increasing the interatomic distance to $r=7\,\mu$m, i.e., decreasing
the interaction strength, we expect that the blockade mechanism becomes ineffective and the contribution of Rydberg
pairs to $f_{e}$ becomes relevant. 
This is
indeed reflected in the fact that the peak of $\rho_{ee;ee}$ in
\fig{fig:twophotonprof}b is orders of magnitude higher than in
\fig{fig:twophotonprof}a.  Here, however, the atoms are
\emph{successively} excited to the Rydberg state by two single-photon
transitions.  Hence, the peak in $\rho_{ee;ee}$ is correctly
reproduced by the rate equation.

\section{Rydberg excitation in large ensembles and comparison with the experiment}\label{exper}
\subsection{Dipole blockade}\label{dblock}
\subsubsection{The density of Rydberg atoms}

\begin{figure}
\includegraphics[width=0.9\columnwidth]{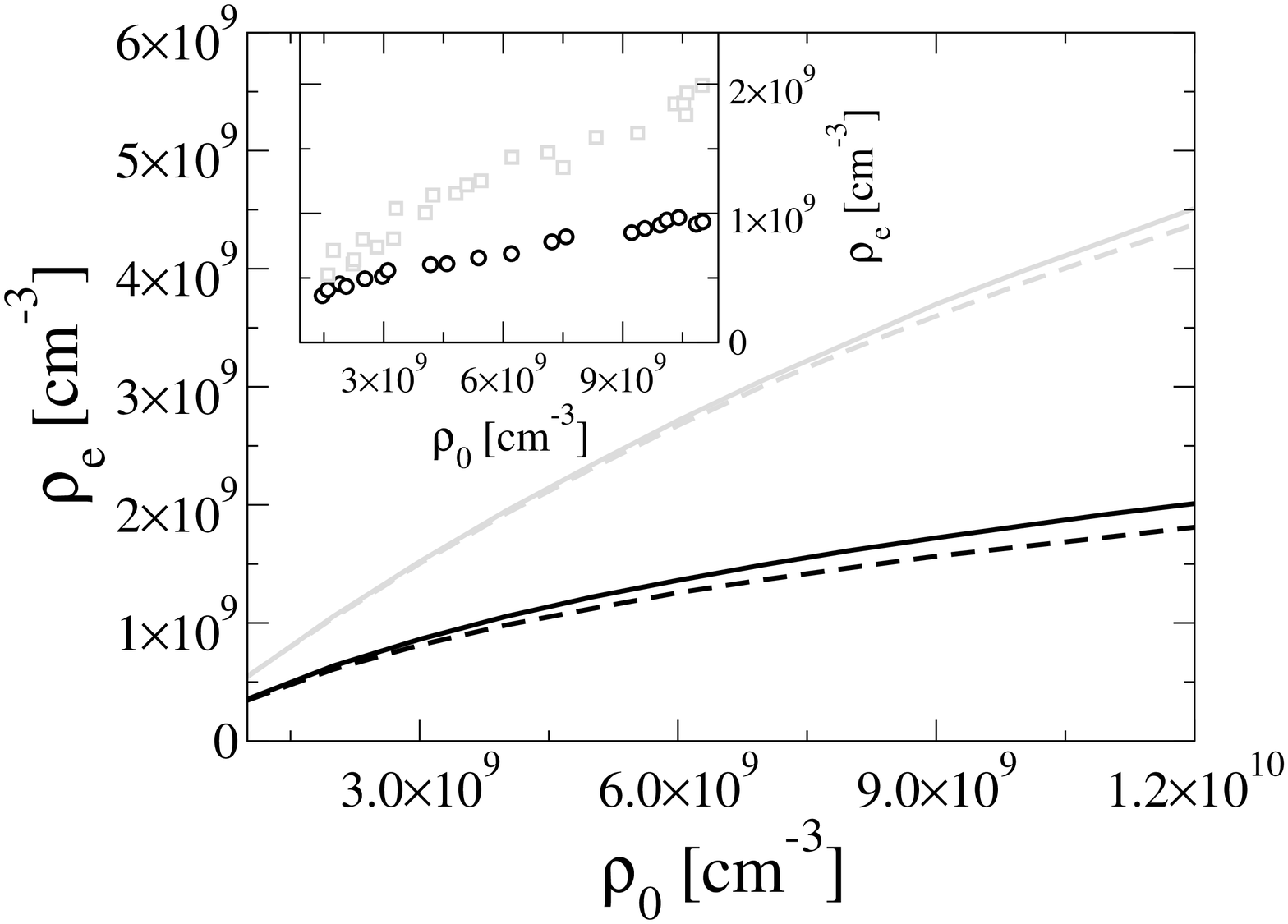}
\caption{Density of Rydberg atoms as a function of the peak density in the MOT
for a pulse length of $\tau=20\,\mu$s for the 82S (black) and the 62S state
(gray) of Rb. Circles:  experimental data taken from \cite{sire+:04}. Lines:
Calculations  using different models for the pair interactions potential:
two-state model of ref. \cite{lita+:05} (solid) and pure van der Waals
interaction from perturbative treatment \cite{sist+:05} (dashed).}
\label{fig:expWM}
\end{figure}

We have calculated the density of Rydberg atoms as a function of the
peak density of a Rb gas in a MOT according to \eq{many_RE} for
excitations to the 62$S$ and 82$S$ state via the two-step excitation scheme
as measured in \cite{sire+:04}.

More specifically, we have determined the Rabi frequency $\Omega$ of
the first excitation step by using the data for the $5S_{1/2}
(F=2) \to 5P_{3/2} (F=3)$ trapping transition of $^{87}$Rb
\cite{st:01} and by taking the intensity of the MOT lasers from the
experiment \cite{we:pc}.  The measurment of $\Omega$ as a function of
the intensity of the MOT lasers using the Autler-Townes splitting of a
Rydberg line \cite{we:pc} is in very good agreement with our result.

To obtain the coupling strength $\omega$ of the Rydberg transition we
have fitted it to the low-intensity measurements in \cite{sire+:04} using our
rate equation and scaled the result to high intensities and/or excitations
to different principal quantum numbers.

Fig. \ref{fig:expWM} shows the results of our calculations and the 
experiment. Although we see a qualitative agreement we predict Rydberg densities 
about twice as large as the measured ones. As the curves for both
principal quantum numbers exhibit the same deviation from the measured data, it
is tempting to scale our results to the experimental points using a common
factor. Note, however, that without other influences in the 
experiment, there is no free parameter in our description that
would justify such a scaling.

In the following we estimate the quantitative influence which several 
effects could have on the results presented.

\subsubsection{The influence of different Rydberg-Rydberg interactions}

The ``exact'' Rydberg-Rydberg interaction may differ from the one we 
have used in our description. To assess the impact of such a 
difference,
we have performed our calculations with the simple two-state
model discussed above (solid lines in \fig{fig:expWM}) and assuming a ``pure''
van der Waals interaction, $-C_6/r^6$, between the Rydberg atoms (dashed lines
in \fig{fig:expWM}).  The interaction coefficients $C_6(n)$ for the latter are
calculated in second-order perturbation theory for $r\to\infty$ and have been
taken from \cite{sist+:05}.  The interaction strength for the $nS$ states
calculated in this way is considerably larger than the one from the two-state
model (e.g., for the $82S$ state the difference in $U(r)$ at $r=10\,\mu$m is
roughly a factor of 2.5 and increases with decreasing $r$). Yet, the 
final results for the Rydberg population
differ only slightly (see \fig{fig:expWM}).  We conclude that
$\rho_{\text e}$ is relatively robust against changes in the interaction
strength. This is due to the fact that the measurement of the Rydberg density as
a function of the ground state density does not probe the exact shape of the
interaction potential but rather the critical distance $r_{\text{c}}$ at which
the energetic shift caused by the interaction becomes larger than half the width
of the spectral line ($\approx 20\,$MHz).  For $U(r)$ determined in perturbation
theory and estimated by the two-state approximation $r_{\text{c}} \approx
8\,\mu$m and $r_{\text{c}}=7\,\mu$m, respectively, for the 82S state, so that
significant differences emerge only for large densities.

\begin{figure}
\centering
 \includegraphics[width=0.9\columnwidth]{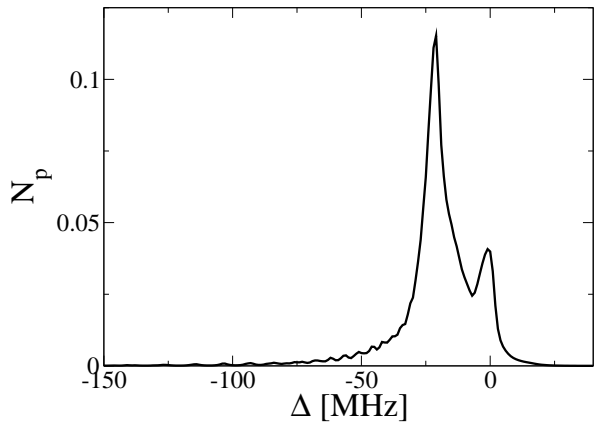}
\caption{Estimated average number $N_p$ of $n=82$ Rydberg pairs excited by
multi-photon transitions as a function of the laser detuning $\Delta$ after
$\tau=20\mu$s for a ground state peak density $\rho_0=10^{10}\,\text{cm}^{-3}$.}
\label{fig:mpp}
\end{figure}

\subsubsection{The influence of ions}

Another effect, so far not accounted for, is the presence of ions.
The excitation pulse length used in \cite{sire+:04} was $20\,\mu$s. For pulse
durations that long, it was shown that a significant amount of Rydberg atoms can
undergo ionizing collisions even for a repulsive Rydberg-Rydberg interaction
\cite{lita+:05,amre+:07}. The presence of ions in the system influences the excitation 
dynamics due to the polarizing effect of the electric field of the ions on the
highly susceptible Rydberg atoms. The Rydberg-ion interaction ($\propto
r^{-4}$), therefore, leads to an additional energetic shift of the Rydberg
levels and, thus, can lead to an enhanced excitation suppression. 

To see if the presence of ions can account for the difference between
our results and the measured data, we have performed calculations in
which we have replaced up to 20\% of the Rydberg atoms by ions.  The
change in the results compared to the situation without ions is
comparable to that of stronger Rydberg-Rydberg interaction discussed
above.  Therefore, ions can be ruled out as a source for the
discrepancy between our and the experimental results.

\subsubsection{The influence of multiphoton transitions}

The excitation line profiles presented in \cite{sire+:04} showed an enormous
broadening for measurements at high densities. In contrast, the line profiles
that we have calculated with the present approach are much narrower, 
in accordance with the simulations reported in ref. \cite{rohe:05}.

The strong line broadening in the experiment could be due to 
non-resonant effects, such as multiphoton transitions, not included in
our rate description (see discussion in section \ref{twoatom}).  To
estimate their possible influence, we have to determine first the
number of Rydberg pairs which could be excited by these transitions.
To this end, we have determined the number of (ground state) atoms
$n_p(r)\Delta r$ which form a pair with a distance between $r$ and
$r+\Delta r$ in the excitation volume, from the pair density
$n_{p}(r)$.  Furthermore, we have calculated the probability
$\rho_{ee;ee}$ for a pair of atoms to be in the Rydberg state after
$\tau=20\mu$s by solving the quantum master equation ($w_{p}^{\text{
ME}}$) and the rate equation ($w_{p}^{\text{ RE}}$) for two atoms as a
function of the laser detuning $\Delta$ and interatomic distance $r$
(c.f.\ \fig{fig:twophotonprof}).  The difference $w_p(r,\Delta) =
w_{p}^{\text{ ME}}(r,\Delta)-w_{p}^{\text{ RE}}(r,\Delta)$ should give
a rough estimate for the probability of a Rydberg pair being excited by
a multi-photon transition.  The average number of such pairs as a
function of $\Delta$ can then be estimated by $N_p(\Delta) = \sum_i
w_p(r_i,\Delta) n_p(r_i)\Delta r$.

Fig. \ref{fig:mpp} shows that for a sample with ground state peak density
$\rho_0=10^{10}\,\text{cm}^{-3}$ our estimate yields a negligible number of
Rydberg pairs excited by multi-photon transitions after $20\mu$s. 
Although these estimates are rather crude, the result shows that 
multi-photon effects are too small to explain the broadening of the 
excitation line profile in the experiment  \cite{sire+:04}.

In summary, the unexplained line broadening and the difference between experiment and 
theory in the Rydberg populations make it likely
that some additional, presently not known process, has contributed
significantly to the results obtained in \cite{sire+:04}.

\subsection{Antiblockade}
\subsubsection{Lattice configurations}\label{antiblock}
The discussion in Sec. \ref{twoatom} has shown that the structure of the single-atom
excitation line strongly influences the excitation dynamics in the interacting
system.  Even on resonance, the Rydberg-Rydberg interaction can cause an
excitation enhancement, if the spectral line exhibits a double peak structure.
This antiblockade occurs whenever the interaction-induced energetic shift
$\Delta_i$ for an atom at position $\mathbf{r}_i$ matches the detuning
$\Delta_{max}$ at which the single-atom excitation probability has its maximum
value.

\begin{figure*}
\centering
\includegraphics[width=0.9\columnwidth]{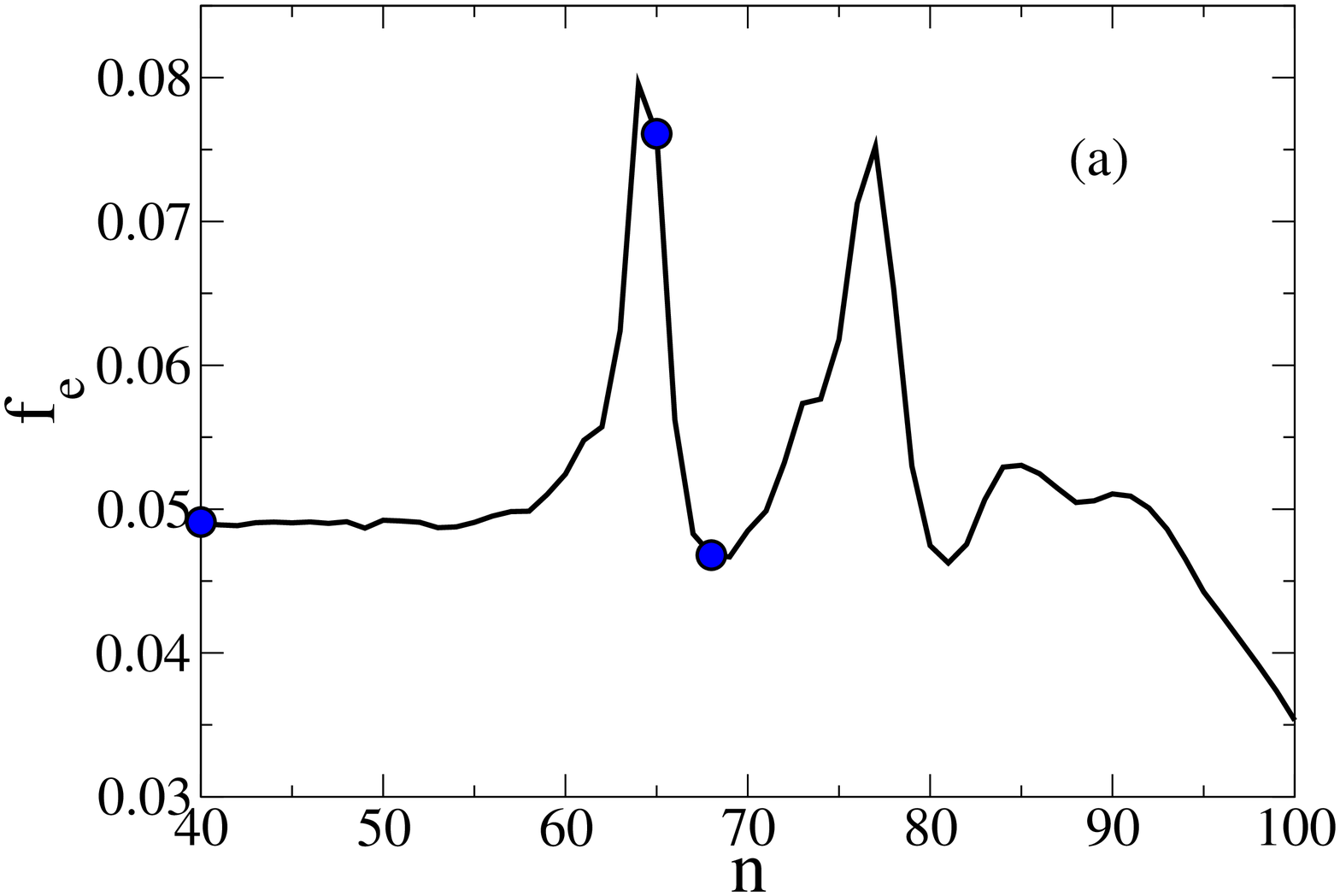} \hfill
\includegraphics[width=0.9\columnwidth]{fig7b}
\caption{(a) Fraction of excited atoms for atoms on a simple cubic lattice with
20\% unoccupied sites as function of the principal quantum number $n$. The lattice
constant is $a=5\,\mu$m, all other parameters are those of 
\fig{fig:populations}b. (b-d) Corresponding number of ``Rydberg clusters''
per lattice site $n_s$ normalized to the number of 1-clusters (i.e. isolated Rydberg atoms) 
$n_1$ as a function of the cluster size $s$ for principal quantum number $n=40$ (b), $n=65$ (c) 
and $n=68$ (d). The shaded areas represent 
predictions from percolation theory \cite{syga+:76} for a system with the same number of isolated Rydberg atoms (1-clusters) per lattice site.}
\label{fig:ablock}
\end{figure*}

In the gas phase, where the mutual atomic distances are broadly
distributed, the antiblockade can hardly be observed by measuring the
fraction of excited atoms $f_e$, as the condition
$\Delta_i=\Delta_{max}$ is only met by relatively few atoms
\cite{atpo+:07a}.  In contrast, if the atoms are regularly arranged in
space, e.g., with the help of an optical lattice produced by
CO$_2$ lasers \cite{frda+:98}, one should clearly observe peaks in
$f_e$ for certain $n$ (see \fig{fig:ablock}a).  The peak
positions can easily be determined by analyzing the geometry of the
underlying lattice.  Moreover, the effect is quite robust against
lattice defects (unoccupied lattice sites) and should therefore be
experimentally realizable. A more detailed discussion can be found in 
 \cite{atpo+:07a}.

The underlying lattice structure allows for a statistical
interpretation of the antiblockade as ``clustering'' of Rydberg atoms.
Using the terminology of percolation theory, we define a cluster of
size $s$ as group of $s$ nearest neighbor sites occupied by Rydberg
atoms.  For negligible Rydberg-Rydberg interaction the excitation of
atoms on a lattice is analogous to the situation encountered in
classical (site-)percolation theory.  This is seen in
\fig{fig:ablock}b, where a histogram of the average number $n_s$ of
$s$-clusters per lattice site as function of the cluster size (normalized to the 
number of 1-clusters, i.e., isolated Rydberg atoms) is shown
for atoms excited to the state $n=40$.  The shaded area represents the
prediction of percolation theory \cite{syga+:76} for the same number of 
isolated Rydberg atoms per site and shows good
agreement with the ``measured'' data.  In the antiblockade regime
($n=65$, \fig{fig:ablock}c) we observe a broadening of the cluster
size distribution and a significant enhancement of larger Rydberg
clusters, while in the blockade regime ($n=68$, \fig{fig:ablock}d) a
quenching of the distribution and an enhancement of the probability to
excite isolated Rydberg atoms is evident.

\subsubsection{Random gases}\label{acs}
Based on the solution of a many-body rate equation using Monte Carlo
sampling, the present approach is particularly well suited to
determine statistical properties of interacting Rydberg gases.

\begin{figure}
\centering
\includegraphics[width=0.9\columnwidth]{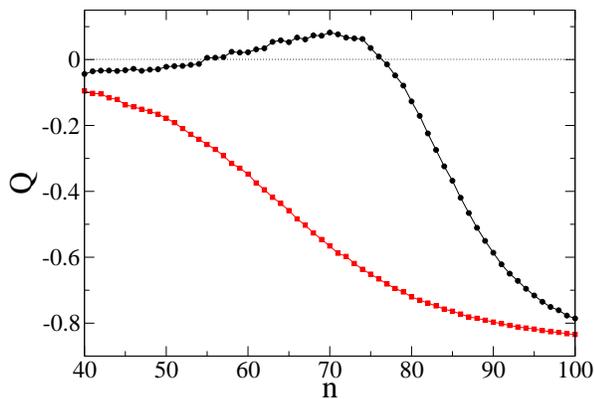}
\caption{Comparison of the $Q$-parameter in the blockade (squares) and
antiblockade (circles) configuration as a function of the principal
quantum number $n$ for a sample with a homogeneous atomic density
$\rho_0=8\times 10^9\,\text{cm}^{-3}$ and for an excitation pulse
length $\tau=2\,\mu$s.  $Q$ was determined by $10^{5}$ successive
measurements of $\langle N_{ryd}\rangle$ and $\langle N^2_{ryd}
\rangle$.  The Rabi frequencies ($\Omega$, $\omega$) are: (4.0, 0.24)
MHz (squares) and (22.1, 0.8) MHz (circles).}
\label{fig:q}
\end{figure}

In \cite{lire+:05} the distribution of the number of Rydberg atoms was
measured as function of the interaction strength.  The 
distributions obtained were quantified by Mandel's $Q$-parameter
\begin{equation}
Q = \frac{\langle N_e^2 \rangle - \langle N_e \rangle^2}{\langle N_e \rangle}
- 1 \; ,
\end{equation}
where $N_e$ is the number of Rydberg atoms and $\langle \dots \rangle$ denotes
the average over the probability distribution. The $Q$-parameter measures the
deviation of a probability distribution from a Poissonian, for which it is
zero, whereas for a super-(sub-)Poissonian it is positive (negative). The
experiment showed a quenching of the Rydberg number distribution, i.e., a
decrease of $Q$, for increasing interaction strength as theoretically 
confirmed \cite{atpo+:06,hero:06}. The differences between the theoretical
calculations ($Q<0$, for all $n$) and the measured values ($Q>0$) can be
attributed to shot-to-shot fluctuations of the number of ground state 
atoms in the experiment
\cite{lire+:07}.

The excitation parameters in \cite{lire+:05} were in the blockade
regime, where the single-atom excitation line exhibits a single peak
at $\Delta=0$.  Therefore, there is a volume (``correlation hole'')
around each Rydberg atom, where the excitation of additional atoms is
strongly suppressed.  On the other hand, in the parameter regime of
the antiblockade, where the excitation line shows a double peak
structure, there is in addition a shell around each Rydberg atom, in
which additional excitations are strongly enhanced.  Thus, the
statistics of the Rydberg excitations should depend on the structure
of the single-atom excitation line and the antiblockade can be
detected indirectly even in the gas phase by measuring the atom
counting statistics.

Figure \ref{fig:q} shows the calculated $Q$-parameter as a function of
the principal quantum number $n$ for the blockade and antiblockade
regime.  In the blockade configuration (squares) one observes a
monotonic decrease of $Q$ with $n$ in accordance with the measurements
in \cite{lire+:05}.  In the antiblockade regime (circles), however,
$Q$ is non-monotonic, i.e., the distribution is slightly broadened,
and the quenching starts at much higher $n$.  Although the broadening
of the distribution may be difficult to observe experimentally, the
difference in the functional form of $Q(n)$ provides a clear
experimental signature in a mesoscopic region of the MOT, where the
atomic density is approximately homogeneous.

\section{Conclusions}\label{final}
We have developed a simple approach, which allows one to describe the
dynamics in ultracold gases, in which Rydberg atoms are excited via a
resonant two-step transition.  Starting from a quantum master
equation, which incorporates the full dynamics of an interacting gas
of three-level atoms, we have derived a many-body rate equation.  It
covers the correlated dynamics of the system, yet, it can easily be
solved by Monte Carlo sampling for a realistically large number of
atoms.

Our approach, valid under well defined conditions typical for
experiments, is based upon two approximations: (i) an adiadabtic
approximation on the \emph{single-atom} level to eliminate the atomic
coherences and (ii) the negligence of multi-photon transitions in the
\emph{interacting system}.
Solving the problem of two interacting atoms exactly with a quantum 
master equation we could show that the approximate solution based on 
the rate equation is in very good agreement with the exact result.

The present approach is capable of reproducing the partial excitation blockade
observed in \cite{sire+:04} qualitatively. 

Qualitatively in accordance with our calculations regarding the 
excitation line shape and the so called Q-parameter are also the 
experimental results of \cite{lire+:05}.

Finally, the careful analysis of the two-step excitation scheme has 
lead to the prediction of an antiblockade effect due to an 
Autler-Townes splitting of the intermediate level probed by the Rydberg transition in the appropriate 
parameter regime. This antiblockade should be directly observable 
for a lattice gas, realized, e.g., with an optical lattice. As we 
have demonstrated, it could also be observed indirectly in the gas 
phase through the atom counting statistics which differs 
qualitatively from its counterpart in the blockade regime.

\appendix*
\section{Expressions for $\rho_{\text{ee}}^{\infty}$ and $\gamma_{\uparrow}$}
The steady state solution of the OBE (\ref{obe}) for the Rydberg population is
\begin{equation}
\rho_{\text{ee}}^{\infty}=
\frac{\Omega^2 \, \left(\Omega^2 + \omega^2 \right)}{\left(\Omega^2 + \omega^2
\right)^2 + 4\Delta^2\,\left(\Gamma^2 + 2\Omega^2 \right)} \, .
\end{equation}
The excitation rate in (\ref{diff_eq}) can be written as 
\begin{equation}
\gamma_{\uparrow}=\frac{2\Gamma\,(\omega\,\Omega)^2\, \left(\Omega^2+\omega^2 \right)}{a_0 + a_2
\Delta^2 + a_4 \Delta^4} \, ,
\end{equation}
where
\begin{subequations}
\begin{eqnarray}
a_0 & = & \left(\Omega^2+\omega^2 \right) \nonumber \\
& & \cdot \left[\left(\omega^2 -2\Omega^2 
\right)^2 +2\Gamma^2\, \left(\Omega^2+\omega^2 \right) \right] \\
a_2 & = & 8\left( \Gamma^4 - 4\Omega^4 \right) \nonumber \\
 & & +4\omega^2\,\left( \Gamma^2
+4\Omega^2\right) + 8 \omega^4 \\
a_4 & = & 32\left(\Gamma^2 + 2\Omega^2 \right) \, .
\end{eqnarray}
\end{subequations}


\begin{thebibliography}{25}
\expandafter\ifx\csname natexlab\endcsname\relax\def\natexlab#1{#1}\fi
\expandafter\ifx\csname bibnamefont\endcsname\relax
  \def\bibnamefont#1{#1}\fi
\expandafter\ifx\csname bibfnamefont\endcsname\relax
  \def\bibfnamefont#1{#1}\fi
\expandafter\ifx\csname citenamefont\endcsname\relax
  \def\citenamefont#1{#1}\fi
\expandafter\ifx\csname url\endcsname\relax
  \def\url#1{\texttt{#1}}\fi
\expandafter\ifx\csname urlprefix\endcsname\relax\def\urlprefix{URL }\fi
\providecommand{\bibinfo}[2]{#2}
\providecommand{\eprint}[2][]{\url{#2}}

\bibitem[{\citenamefont{Gallagher}(1994)}]{ga:94}
\bibinfo{author}{\bibfnamefont{T.~F.} \bibnamefont{Gallagher}},
  \emph{\bibinfo{title}{Rydberg atoms}} (\bibinfo{publisher}{Cambridge
  University Press}, \bibinfo{year}{1994}).

\bibitem[{\citenamefont{Anderson et~al.}(1998)\citenamefont{Anderson, Veale,
  and Gallagher}}]{anve+:98}
\bibinfo{author}{\bibfnamefont{W.~R.} \bibnamefont{Anderson}},
  \bibinfo{author}{\bibfnamefont{J.~R.} \bibnamefont{Veale}}, \bibnamefont{and}
  \bibinfo{author}{\bibfnamefont{T.~F.} \bibnamefont{Gallagher}},
  \bibinfo{journal}{Phys. Rev. Lett.} \textbf{\bibinfo{volume}{80}},
  \bibinfo{pages}{249} (\bibinfo{year}{1998}).

\bibitem[{\citenamefont{Mourachko et~al.}(1998)\citenamefont{Mourachko,
  Comparat, de~Tomasi, Fioretti, Nosbaum, Akulin, and Pillet}}]{moco+:98}
\bibinfo{author}{\bibfnamefont{I.}~\bibnamefont{Mourachko}},
  \bibinfo{author}{\bibfnamefont{D.}~\bibnamefont{Comparat}},
  \bibinfo{author}{\bibfnamefont{F.}~\bibnamefont{de~Tomasi}},
  \bibinfo{author}{\bibfnamefont{A.}~\bibnamefont{Fioretti}},
  \bibinfo{author}{\bibfnamefont{P.}~\bibnamefont{Nosbaum}},
  \bibinfo{author}{\bibfnamefont{V.~M.} \bibnamefont{Akulin}},
  \bibnamefont{and} \bibinfo{author}{\bibfnamefont{P.}~\bibnamefont{Pillet}},
  \bibinfo{journal}{Phys. Rev. Lett.} \textbf{\bibinfo{volume}{80}},
  \bibinfo{pages}{253} (\bibinfo{year}{1998}).

\bibitem[{\citenamefont{Lukin et~al.}(2001)\citenamefont{Lukin, Fleischhauer,
  C\^ot\'e, Duan, Jaksch, Cirac, and Zoller}}]{lufl+:01}
\bibinfo{author}{\bibfnamefont{M.~D.} \bibnamefont{Lukin}},
  \bibinfo{author}{\bibfnamefont{M.}~\bibnamefont{Fleischhauer}},
  \bibinfo{author}{\bibfnamefont{R.}~\bibnamefont{C\^ot\'e}},
  \bibinfo{author}{\bibfnamefont{L.~M.} \bibnamefont{Duan}},
  \bibinfo{author}{\bibfnamefont{D.}~\bibnamefont{Jaksch}},
  \bibinfo{author}{\bibfnamefont{J.~I.} \bibnamefont{Cirac}}, \bibnamefont{and}
  \bibinfo{author}{\bibfnamefont{P.}~\bibnamefont{Zoller}},
  \bibinfo{journal}{Phys. Rev. Lett.} \textbf{\bibinfo{volume}{87}},
  \bibinfo{pages}{037901} (\bibinfo{year}{2001}).

\bibitem[{\citenamefont{Bouchoule and M\o{}lmer}(2002)}]{bomo:02}
\bibinfo{author}{\bibfnamefont{I.}~\bibnamefont{Bouchoule}} \bibnamefont{and}
  \bibinfo{author}{\bibfnamefont{K.}~\bibnamefont{M\o{}lmer}},
  \bibinfo{journal}{Phys. Rev. A} \textbf{\bibinfo{volume}{65}},
  \bibinfo{pages}{041803} (\bibinfo{year}{2002}).

\bibitem[{\citenamefont{Saffman and Walker}(2002)}]{sawa:02}
\bibinfo{author}{\bibfnamefont{M.}~\bibnamefont{Saffman}} \bibnamefont{and}
  \bibinfo{author}{\bibfnamefont{T.~G.} \bibnamefont{Walker}},
  \bibinfo{journal}{Phys. Rev. A} \textbf{\bibinfo{volume}{66}},
  \bibinfo{pages}{065403} (\bibinfo{year}{2002}).

\bibitem[{\citenamefont{Tong et~al.}(2004)\citenamefont{Tong, Farooqi,
  Stanojevic, Krishnan, Zhang, C\^ot\'e, Eyler, and Gould}}]{tofa+:04}
\bibinfo{author}{\bibfnamefont{D.}~\bibnamefont{Tong}},
  \bibinfo{author}{\bibfnamefont{S.~M.} \bibnamefont{Farooqi}},
  \bibinfo{author}{\bibfnamefont{J.}~\bibnamefont{Stanojevic}},
  \bibinfo{author}{\bibfnamefont{S.}~\bibnamefont{Krishnan}},
  \bibinfo{author}{\bibfnamefont{Y.~P.} \bibnamefont{Zhang}},
  \bibinfo{author}{\bibfnamefont{R.}~\bibnamefont{C\^ot\'e}},
  \bibinfo{author}{\bibfnamefont{E.~E.} \bibnamefont{Eyler}}, \bibnamefont{and}
  \bibinfo{author}{\bibfnamefont{P.~L.} \bibnamefont{Gould}},
  \bibinfo{journal}{Phys. Rev. Lett.} \textbf{\bibinfo{volume}{93}},
  \bibinfo{pages}{063001} (\bibinfo{year}{2004}).

\bibitem[{\citenamefont{Singer et~al.}(2004)\citenamefont{Singer, Reetz-Lamour,
  Amthor, Marcassa, and Weidem\"uller}}]{sire+:04}
\bibinfo{author}{\bibfnamefont{K.}~\bibnamefont{Singer}},
  \bibinfo{author}{\bibfnamefont{M.}~\bibnamefont{Reetz-Lamour}},
  \bibinfo{author}{\bibfnamefont{T.}~\bibnamefont{Amthor}},
  \bibinfo{author}{\bibfnamefont{L.~G.} \bibnamefont{Marcassa}},
  \bibnamefont{and}
  \bibinfo{author}{\bibfnamefont{M.}~\bibnamefont{Weidem\"uller}},
  \bibinfo{journal}{Phys. Rev. Lett.} \textbf{\bibinfo{volume}{93}},
  \bibinfo{pages}{163001} (\bibinfo{year}{2004}).

\bibitem[{\citenamefont{Vogt et~al.}(2006)\citenamefont{Vogt, Viteau, Zhao,
  Chotia, Comparat, and Pillet}}]{vovi+:06}
\bibinfo{author}{\bibfnamefont{T.}~\bibnamefont{Vogt}},
  \bibinfo{author}{\bibfnamefont{M.}~\bibnamefont{Viteau}},
  \bibinfo{author}{\bibfnamefont{J.}~\bibnamefont{Zhao}},
  \bibinfo{author}{\bibfnamefont{A.}~\bibnamefont{Chotia}},
  \bibinfo{author}{\bibfnamefont{D.}~\bibnamefont{Comparat}}, \bibnamefont{and}
  \bibinfo{author}{\bibfnamefont{P.}~\bibnamefont{Pillet}},
  \bibinfo{journal}{Phys. Rev. Lett.} \textbf{\bibinfo{volume}{97}},
  \bibinfo{eid}{083003} (\bibinfo{year}{2006}).

\bibitem[{\citenamefont{Cubel-Liebisch
  et~al.}(2005)\citenamefont{Cubel-Liebisch, Reinhard, Berman, and
  Raithel}}]{lire+:05}
\bibinfo{author}{\bibfnamefont{T.}~\bibnamefont{Cubel-Liebisch}},
  \bibinfo{author}{\bibfnamefont{A.}~\bibnamefont{Reinhard}},
  \bibinfo{author}{\bibfnamefont{P.~R.} \bibnamefont{Berman}},
  \bibnamefont{and} \bibinfo{author}{\bibfnamefont{G.}~\bibnamefont{Raithel}},
  \bibinfo{journal}{Phys. Rev. Lett.} \textbf{\bibinfo{volume}{95}},
  \bibinfo{pages}{253002} (\bibinfo{year}{2005}).

\bibitem[{\citenamefont{Ates et~al.}(2006)\citenamefont{Ates, Pohl, Pattard,
  and Rost}}]{atpo+:06}
\bibinfo{author}{\bibfnamefont{C.}~\bibnamefont{Ates}},
  \bibinfo{author}{\bibfnamefont{T.}~\bibnamefont{Pohl}},
  \bibinfo{author}{\bibfnamefont{T.}~\bibnamefont{Pattard}}, \bibnamefont{and}
  \bibinfo{author}{\bibfnamefont{J.~M.} \bibnamefont{Rost}},
  \bibinfo{journal}{J. Phys. B} \textbf{\bibinfo{volume}{39}},
  \bibinfo{pages}{L233} (\bibinfo{year}{2006}).

\bibitem[{\citenamefont{Hern\'andez and Robicheaux}(2006)}]{hero:06}
\bibinfo{author}{\bibfnamefont{J.~V.} \bibnamefont{Hern\'andez}}
  \bibnamefont{and}
  \bibinfo{author}{\bibfnamefont{F.}~\bibnamefont{Robicheaux}},
  \bibinfo{journal}{J. Phys. B} \textbf{\bibinfo{volume}{39}},
  \bibinfo{pages}{4883} (\bibinfo{year}{2006}).

\bibitem[{\citenamefont{Robicheaux and Hern\'andez}(2005)}]{rohe:05}
\bibinfo{author}{\bibfnamefont{F.}~\bibnamefont{Robicheaux}} \bibnamefont{and}
  \bibinfo{author}{\bibfnamefont{J.~V.} \bibnamefont{Hern\'andez}},
  \bibinfo{journal}{Phys. Rev. A} \textbf{\bibinfo{volume}{72}},
  \bibinfo{pages}{063403} (\bibinfo{year}{2005}).

\bibitem[{\citenamefont{Cohen-Tannoudji
  et~al.}(1992)\citenamefont{Cohen-Tannoudji, Dupont-Roc, and
  Grynberg}}]{codu+:92}
\bibinfo{author}{\bibfnamefont{C.}~\bibnamefont{Cohen-Tannoudji}},
  \bibinfo{author}{\bibfnamefont{J.}~\bibnamefont{Dupont-Roc}},
  \bibnamefont{and} \bibinfo{author}{\bibfnamefont{G.}~\bibnamefont{Grynberg}},
  \emph{\bibinfo{title}{Atom-photon interactions: basic processes and
  applications}} (\bibinfo{publisher}{John Wiley \& Sons},
  \bibinfo{year}{1992}).

\bibitem[{\citenamefont{Ates et~al.}(2007)\citenamefont{Ates, Pohl, Pattard,
  and Rost}}]{atpo+:07a}
\bibinfo{author}{\bibfnamefont{C.}~\bibnamefont{Ates}},
  \bibinfo{author}{\bibfnamefont{T.}~\bibnamefont{Pohl}},
  \bibinfo{author}{\bibfnamefont{T.}~\bibnamefont{Pattard}}, \bibnamefont{and}
  \bibinfo{author}{\bibfnamefont{J.~M.} \bibnamefont{Rost}},
  \bibinfo{journal}{Phys. Rev. Lett.} \textbf{\bibinfo{volume}{98}},
  \bibinfo{pages}{023002} (\bibinfo{year}{2007}).

\bibitem[{\citenamefont{Singer et~al.}(2005)\citenamefont{Singer, Stanojevic,
  Weidem\"uller, and C\^ot\'e}}]{sist+:05}
\bibinfo{author}{\bibfnamefont{K.}~\bibnamefont{Singer}},
  \bibinfo{author}{\bibfnamefont{J.}~\bibnamefont{Stanojevic}},
  \bibinfo{author}{\bibfnamefont{M.}~\bibnamefont{Weidem\"uller}},
  \bibnamefont{and} \bibinfo{author}{\bibfnamefont{R.}~\bibnamefont{C\^ot\'e}},
  \bibinfo{journal}{J. Phys. B} \textbf{\bibinfo{volume}{38}},
  \bibinfo{pages}{S295} (\bibinfo{year}{2005}).

\bibitem[{\citenamefont{Reinhard et~al.}(2007)\citenamefont{Reinhard, Liebisch,
  Knuffman, and Raithel}}]{recu+:07}
\bibinfo{author}{\bibfnamefont{A.}~\bibnamefont{Reinhard}},
  \bibinfo{author}{\bibfnamefont{T.~C.} \bibnamefont{Liebisch}},
  \bibinfo{author}{\bibfnamefont{B.}~\bibnamefont{Knuffman}}, \bibnamefont{and}
  \bibinfo{author}{\bibfnamefont{G.}~\bibnamefont{Raithel}},
  \bibinfo{journal}{Phys. Rev. A} \textbf{\bibinfo{volume}{75}},
  \bibinfo{eid}{032712} (\bibinfo{year}{2007}).

\bibitem[{\citenamefont{Schwettmann et~al.}(2006)\citenamefont{Schwettmann,
  Crawford, Overstreet, and Shaffer}}]{sccr+:06}
\bibinfo{author}{\bibfnamefont{A.}~\bibnamefont{Schwettmann}},
  \bibinfo{author}{\bibfnamefont{J.}~\bibnamefont{Crawford}},
  \bibinfo{author}{\bibfnamefont{K.~R.} \bibnamefont{Overstreet}},
  \bibnamefont{and} \bibinfo{author}{\bibfnamefont{J.~P.}
  \bibnamefont{Shaffer}}, \bibinfo{journal}{Phys. Rev. A}
  \textbf{\bibinfo{volume}{74}}, \bibinfo{eid}{020701} (\bibinfo{year}{2006}).

\bibitem[{\citenamefont{Li et~al.}(2005)\citenamefont{Li, Tanner, and
  Gallagher}}]{lita+:05}
\bibinfo{author}{\bibfnamefont{W.}~\bibnamefont{Li}},
  \bibinfo{author}{\bibfnamefont{P.~J.} \bibnamefont{Tanner}},
  \bibnamefont{and} \bibinfo{author}{\bibfnamefont{T.~F.}
  \bibnamefont{Gallagher}}, \bibinfo{journal}{Phys. Rev. Lett.}
  \textbf{\bibinfo{volume}{94}}, \bibinfo{pages}{173001}
  (\bibinfo{year}{2005}).

\bibitem[{\citenamefont{Steck}(2003)}]{st:01}
\bibinfo{author}{\bibfnamefont{D.~A.} \bibnamefont{Steck}},
  \emph{\bibinfo{title}{Rubidium 87 {D} line data}} (\bibinfo{year}{2003}),
  \urlprefix\url{http://steck.us/alkalidata}.

\bibitem[{\citenamefont{Weidem\"uller}()}]{we:pc}
\bibinfo{author}{\bibfnamefont{M.}~\bibnamefont{Weidem\"uller}},
  \emph{\bibinfo{title}{private communication}}.

\bibitem[{\citenamefont{Amthor et~al.}(2007)\citenamefont{Amthor, Reetz-Lamour,
  Westermann, Denskat, and Weidemuller}}]{amre+:07}
\bibinfo{author}{\bibfnamefont{T.}~\bibnamefont{Amthor}},
  \bibinfo{author}{\bibfnamefont{M.}~\bibnamefont{Reetz-Lamour}},
  \bibinfo{author}{\bibfnamefont{S.}~\bibnamefont{Westermann}},
  \bibinfo{author}{\bibfnamefont{J.}~\bibnamefont{Denskat}}, \bibnamefont{and}
  \bibinfo{author}{\bibfnamefont{M.}~\bibnamefont{Weidemuller}},
  \bibinfo{journal}{Phys. Rev. Lett.} \textbf{\bibinfo{volume}{98}},
  \bibinfo{eid}{023004} (\bibinfo{year}{2007}).

\bibitem[{\citenamefont{Friebel et~al.}(1998)\citenamefont{Friebel,
  D\char39{}Andrea, Walz, Weitz, and H\"ansch}}]{frda+:98}
\bibinfo{author}{\bibfnamefont{S.}~\bibnamefont{Friebel}},
  \bibinfo{author}{\bibfnamefont{C.}~\bibnamefont{D\char39{}Andrea}},
  \bibinfo{author}{\bibfnamefont{J.}~\bibnamefont{Walz}},
  \bibinfo{author}{\bibfnamefont{M.}~\bibnamefont{Weitz}}, \bibnamefont{and}
  \bibinfo{author}{\bibfnamefont{T.~W.} \bibnamefont{H\"ansch}},
  \bibinfo{journal}{Phys. Rev. A} \textbf{\bibinfo{volume}{57}},
  \bibinfo{pages}{R20} (\bibinfo{year}{1998}).

\bibitem[{\citenamefont{Sykes et~al.}(1976)\citenamefont{Sykes, Gaunt, and
  Glen}}]{syga+:76}
\bibinfo{author}{\bibfnamefont{M.~F.} \bibnamefont{Sykes}},
  \bibinfo{author}{\bibfnamefont{D.~S.} \bibnamefont{Gaunt}}, \bibnamefont{and}
  \bibinfo{author}{\bibfnamefont{M.}~\bibnamefont{Glen}}, \bibinfo{journal}{J.
  Phys. A} \textbf{\bibinfo{volume}{9}}, \bibinfo{pages}{1705}
  (\bibinfo{year}{1976}).

\bibitem[{\citenamefont{Cubel-Liebisch
  et~al.}(2007)\citenamefont{Cubel-Liebisch, Reinhard, Berman, and
  Raithel}}]{lire+:07}
\bibinfo{author}{\bibfnamefont{T.}~\bibnamefont{Cubel-Liebisch}},
  \bibinfo{author}{\bibfnamefont{A.}~\bibnamefont{Reinhard}},
  \bibinfo{author}{\bibfnamefont{P.~R.} \bibnamefont{Berman}},
  \bibnamefont{and} \bibinfo{author}{\bibfnamefont{G.}~\bibnamefont{Raithel}},
  \bibinfo{journal}{Phys. Rev. Lett.} \textbf{\bibinfo{volume}{98}},
  \bibinfo{eid}{109903} (\bibinfo{year}{2007}).

\end{thebibliography}

\end{document}